\documentclass[doublecol]{epl2} 
\usepackage{graphicx}
\usepackage{amsmath}

\graphicspath{{./FIGS/}}

\usepackage{subfig}

\renewcommand{\dim}{\dint}
\newcommand{\dint}{1} % internal dimension manifold
\newcommand{\Sava}{{\cal S}} % avalanche total size
\newcommand{\lava}{{\ell}} % avalanche total lateral size
\newcommand{\Savax}{\Sava_{x}} % avalanche total size of any point x
\newcommand{\Savaphi}{\Sava_{\phi}} % avalanche total size submanifold of a typical point in the avalanche
\newcommand{\Savatip}{\Sava_{1}} % avalanche total size of x=1
\newcommand{\Savai}{\Sava_{\tt i}} % avalanche total size of x
\newcommand{\Savaxaverage}{\langle \Savax\rangle} % avalanche total size submanifold
\newcommand{\Savamax}{\Sava_{\rm max}} % avalanche total size cutoff
\newcommand{\lcross}{L_{m}} % avalanche lateral size cutoff
\newcommand{\Scross}{{\Sava}_{m}^{\rm cross}} % avalanche lateral size cutoff
\newcommand{\tauS}{\tau} % decay exponent total size global
\newcommand{\tauphi}{{{\tau}_{\phi}}} % decay exponent total size submanifold
\newcommand{\msq}{m^2} % parabola curvature
\newcommand{\cel}{c} %Elasticity of the interface
\newcommand{\msqc}{m^2} %"order" parameter
\newcommand{\zetadep}{\zeta} % depinning roughness
\newcommand{\Lmax}{L} %System size in 1d
\newcommand{\PS}{P(\Sava)} %Distribution
\title{Avalanches in Tip-Driven Interfaces in Random Media}
\author{L.E. Arag\'on \and A.B. Kolton \inst{1} \and P. Le Doussal   \and K.J. Wiese \inst{2} \and E.A. Jagla \inst{1}}
\shortauthor{L. E. Arag\'on \etal}

\institute{                    
  \inst{1} CONICET - Centro At\'omico Bariloche and Instituto Balseiro (UNCu) - (8400) Bariloche, Argentina\\
  \inst{2} CNRS - Laboratoire de Physique Th\'eorique de l'Ecole Normale Superieure - 24 rue Lhomond, 75005 Paris, France 
}
\abstract{We analyse by numerical simulations and scaling arguments 
the avalanche statistics of 1-dimensional elastic interfaces in random media driven 
 at a single point.  Both global and local avalanche sizes are  power-law distributed, with  universal exponents
given  by the depinning roughness exponent $\zeta$ and the interface dimension  $d$, and distinct from  their values in the uniformly driven case.  A crossover appears  between uniformly driven behaviour for small avalanches, and point driven behaviour for large avalanches. 
The scale of the crossover is controlled by the ratio between the stiffness of the pulling spring and the elasticity of the interface; 
it is visible both in the global and local avalanche-size distributions, as in the average spatial avalanche  shape. 
Our results are relevant to model experiments involving locally driven elastic manifolds at 
low temperatures, such as magnetic domain walls or vortex lines in superconductors. }

\begin{document}

\maketitle

\section{Introduction}
An elastic interface  in a  random potential  
is a paradigmatic model %that has many different experimental realizations, 
for the depinning of many apparently unrelated complex 
nonlinear systems. 
Typical examples are weakly pinned vortex lattices in superconductors driven by a super-current, 
charge-density waves %in inhomogeneous media 
driven by an external electric field %\cite{Fisher_1985} 
and
stick-slip motion of seismic faults driven by tectonic loading \cite{BlatterFeigelmanGeshkenbeinLarkinVinokur1994,DSFisher1998,LeDoussal2010Book}.

\begin{figure}
 %\onefigure{Fig1}
\includegraphics[width=0.999\linewidth,page=1]{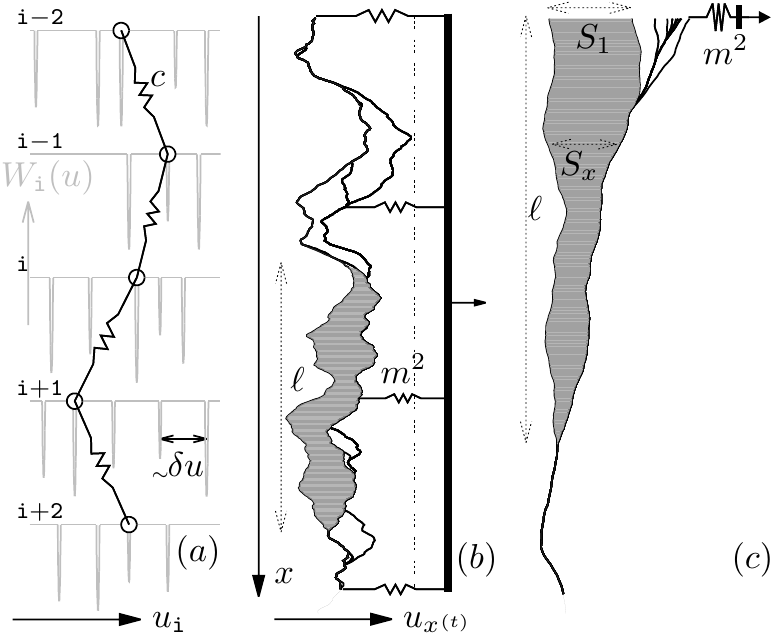}
\caption{
\small (a) A  discrete elastic line with stiffness $\cel$.
The pinning potential $W_{\tt i}$ shown in gray consists of narrow wells.
(b,c) A sequence of configurations for a $1$D interface on a disordered potential, for uniform  (b) 
and localized driving (c). For the latter  
 the tip is pulled with a spring of stiffness $\msq$.
In (c) an average parabolic profile has been subtracted from the picture.
In both cases, the shaded regions represent an avalanche of length $\lava$ and size $\Sava=\int_{\lava} {\rm d} x\, \Savax$. 
}
\label{fig_sketch}
\end{figure}

Usually homogeneous 
driving is  considered, either by applying a constant force on each point of the interface, 
or by attaching a spring to each point and moving its other end at a fixed velocity. 
In both cases, the dynamics proceeds by a sequence of avalanches of size $\Sava$, 
 power-law distributed with  exponent $\tauS$
and   large-size cutoff $\Savamax$,
\begin{equation}
P(\Sava)\sim \Sava^{-\tauS}{\tt g}(\Sava/\Savamax),
 \label{l_PS}
 \end{equation}
where ${\tt g}(x)$ decays rapidly to zero as $x\gg 1$.
The value of $\tauS$ depends on the interface dimension $d$ and its roughness exponent $\zeta$. 
It is robust against many details of the system, 
for example wether it is driven at  constant force or at constant velocity.
For harmonic elastic interfaces uniformly driven, $\tauS$ is around 1.11 in $1$ dimension, and 1.27 in $2$ dimensions.
Avalanche exponents and observables have been recently calculated 
beyond mean-field using renormalization-group methods \cite{LeDoussalWiese2008c,LeDoussalWiese2012a,DobrinevskiLeDoussalWiese2014a}, 
mostly for homogeneous driving.

When driven homogeneously at constant velocity (Fig. \ref{fig_sketch}b),
for sufficiently large systems, 
the external springs of stiffness $\msq$ thus not only drive the system, but also provide a ``mass'' $m^2$ which 
cuts fluctuations beyond a scale $L_{m}\simeq 1/m$.
Therefore the cutoff scales as $\Savamax \sim m^{{-(1+\zeta)}}$ 
which means that the system displays a  power law (and thus a ``critical'' state) only  in the limit $\msqc\to 0$ \cite{Rosso_2009}. %VER DE PONER LAS REFS de TAU125. %LL 
In consequence, it may not be the best model
to describe experiments in which a 
  {\em self-organized critical} state is present, and
the system size is the only  large-scale cutoff.
This   includes most notably  earthquakes, whose size is only limited  by the extension of the tectonic
plates  \cite{Scholz_2002}. 

Here we study the evolution of elastic interfaces  under {\em inhomogeneous} driving. 
To do so, we pull at the tip of a finite one-dimensional string of length $L$ through a single spring of stiffness $\msqc$ 
(Fig.~\ref{fig_sketch}c). 
Once a stationary state is reached,  
the  string has on average to move at the same  velocity as the driving point, 
which  means that  
as   $L$ is increased progressively larger avalanches  occur, independent of the value of $\msqc$. 
Hence this model displays criticality for any value of $\msqc$,
%and the large-scale cutoff is the system size $L$, 
in contrast with the normal homogeneous case, that requires $\msqc\to0$. 

From an experimental point of view, localized  driving appears in many systems:
In the seismic context, the relative motion of plates in subduction zones is mainly driven by the movement of the plates at  regions remote from the seismogenic zone,
i.e.\ from a border of the system~\cite{Rubinstein_2011}.  
Another realization are
vortex lines trapped along a twin boundary plane in a superconductor~\cite{Shapira_2015}.  This effectively 1-dimensional 
elastic interface  can be manipulated through 
a scanning microscope  in a way similar
to our driving at a single point.
Another experiment is a     sandpile, where  sand grains  are  deposited at a given position, leading to a sandpile with a  stationary slope,  evolving through a sequence of avalanches \cite{Frette_1996}.
Self-organized critical systems are further studied 
with cellular automata  like the Oslo model\cite{Christensen_1996}, which is  driven at the boundary. 
In Ref~\cite{Paczuski_1996} the Oslo model for sandpiles was mapped onto a discrete model of an elastic interface pulled at one end 
and it was proposed that it belongs to the same universality class as the inhomogeneous Burridge-Knopoff model \cite{deSousaVieira_1992}
for earthquakes. 
A stochastic version of this friction model was then mapped back to the Oslo model~\cite{Chianca_2009}.
Further exact connections between Manna sandpiles and disordered elastic interfaces have been demonstrated recently~\cite{LeDoussalWiese2014a,Wiese_2015}.
In this Letter we study a tip-driven elastic line as a model system for these phenomena and analyze the avalanche dynamics.

\section{Model and Methods}
\label{l_model}
We model an elastic interface driven on a disordered substrate at zero temperature 
as a discrete string composed of $\Lmax$ particles whose positions 
$u_{\tt i}$ are coupled by elastic 
springs with   Hooke constant $c$, as depicted in Fig.~\ref{fig_sketch}(a). 
We consider an infinitely long disordered medium in the direction of displacements and 
open boundary conditions in the perpendicular direction.   
The pinning to the substrate is modeled by a set of narrow 
potential wells, separated a typical distance $\delta u$. 
Each well is characterized by the maximum 
force it can sustain, which is a positive bounded random value, uncorrelated 
for different pinning wells 
(see \cite{Jagla-Landes-Rosso_PRL_2014} for   details).

We consider two driving protocols: (i) {\em Homogenous driving} with a driving force $\sigma_{\tt i}(t)=\msqc[w(t) - u_{\tt i}(t)]$ for each of the $L$ positions $u_i(t)$, see Fig.~\ref{fig_sketch}b. 
(ii) {\em Inhomogenous driving} %({\em tip driving})
 only at the tip of the string, $\sigma_{\tt 1}(t)=\msqc[w(t) - u_{\tt 1}(t)]$ and all other $\sigma_{\tt i}=0$, see  Fig.~\ref{fig_sketch}c. 
All springs have the  same Hooke constant $m^2$. 
While the homogenous case is well studied, here we focus on the tip-driven case, and compare it to the former.
The system is in a metastable static configuration, 
with all particles sitting on individual pinning wells
as long as the elastic forces on every particle are lower than the pinning forces $f^{\rm pin}_{\tt i}$ 
in a given   configuration of the string. 
When this condition breaks down, an avalanche occurs, see the shaded areas in Figs.~\ref{fig_sketch} (b)-(c), 
until equilibrium is restored in a new static configuration compatible with the partially new set of pinning 
thresholds and elastic forces. We consider   quasi-static driving by fixing  
the position of the driving springs  during the avalanche, which is thus the fastest process. 

An avalanche is characterized by its total spatial extension $\lava$ (number of sites involved) and its size $\Sava$. The latter is the sum of all displacements $\Savai$ during the rearrangement process, namely, $\Sava=\sum_{\tt i}\Savai$.  We also study the 
displacement  $\Sava_{1}$ of the tip of the interface, as depicted in Fig.~\ref{fig_sketch}~(c).   
We analyze the steady state, 
where the sequence of metastable configurations   
advances in the direction of the driving and is unique for a given realization of thresholds~\cite{Middleton1992}. 

We present results obtained  
by choosing an exponential distribution for the separation $\delta u$ between the pinning wells, % centered at $0.1$ 
and a Gaussian distribution for the threshold forces $f_{x}^{\rm pin}$~\cite{Jagla-Landes-Rosso_PRL_2014}.
We consider the position of the driving spring endpoint to increase linearly with time, $w(t)=Vt$, with $V=1$.
Results depend on the ratio   $\msqc/c$ of elastic constants.
We set $\cel=1$, and give the results directly in terms of $\msqc$. 

\section{Results}
\begin{figure}
\centering
\includegraphics[width=0.95\linewidth,page=1]{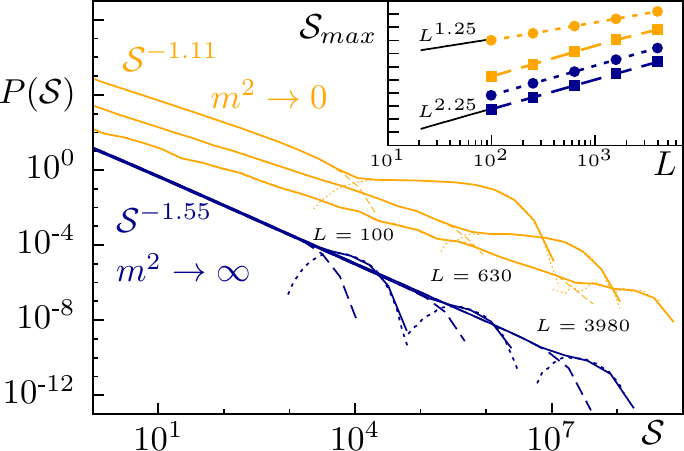}
 \caption{\small (Color online) Distribution of avalanche sizes $\PS$ for different system sizes $\Lmax$ when driving the system by one end. 
 Results are for a driving spring of stiffness $\msq=1$ (dark-blue) and $\msq=10^{-4}$ (light-yellow). 
 In dotted and dashed lines we show the contributions from avalanches that reach or do not reach the full system size. 
 %In the inset we present the 
 Inset: scaling of the cutoff $\Savamax$ with systme size. 
 Squares (circles) ara data from avalanches with $\lava<\Lmax$ ($\lava=\Lmax$).
}
\label{fig_PS}
\end{figure}

We start by computing the global avalanche-size distribution $P(\Sava)$ for tip-driven interfaces, see Fig.~\ref{fig_PS}.
This is equivalent to the distribution of the shaded areas depicted in Fig.~\ref{fig_sketch} (c), generated by the movement of the tip.  
First we analyze the limit $\msqc \to \infty$, when the driving spring is much harder than the inter-particle springs. 
We observe that $P(\Sava)$ decays as a power law 
over more than 6 orders of magnitude for the largest system, with a well-defined exponent 
$\tau \approx 1.55$ and a cutoff $\Savamax$ in Eq.~(\ref{l_PS}). 
This avalanche exponent is compatible with the formula
\begin{equation}
\tau=2-\frac{1}{\dim +\zeta}\ ,
\label{tau1}
\end{equation}
where $\zeta=1.25$ is the well-known  
roughness exponent of  metastable configurations at depinning, taken from measurements for a 
uniformly driven interface~\cite{Ferrero_2013,LeDoussal_2002,Rosso_2001,Chauve_2001}
 (such as those depicted in Fig.~\ref{fig_sketch} (b). 
Interestingly, the same roughness   exponent appears here in the  tip-driven case. It  can be obtained  by subtracting from the metastable configurations the average parabolic profile\footnote{The so-called statistical tilt symmetry (STS) allows to justify that this subtraction yields the standard roughness.}, 
and then calculating the structure factor which 
scales as $|S(q)|^2\sim|q|^{-1-2\zetadep}$. 
The subtracted average  profile is due to the localized driving, and    balances   elastic 
and pinning forces, 
$ \partial^2_x u \sim f_{x}^{\rm pin}$. 
An argument in favor of the scaling relation (\ref{tau1}) was given in Ref.~\cite{Paczuski_1996}. 
We have verified that this formula also holds for tip-driven static avalanches 
connecting stable equilibrium states where $\zeta=2/3$~\cite{Huse_1985,Kardar_1985}.

The avalanche exponent of Eq.~(\ref{tau1}) is   distinct  from the value 
\begin{equation}
\tau=2-\frac{2}{\dim+\zeta}, 
\label{tau2}
\end{equation}
valid for  uniformly driven interfaces, see Fig.~\ref{fig_sketch} (b). 
Using $\zeta=1.25$ %and $\dim=1$ 
one obtains $\tau\approx 1.11$. 
Interestingly, as shown in Fig.~\ref{fig_PS}, the exponent 
$\tau \approx 1.11$ is recovered in the limit  $\msqc\to 0$ of the tip-driven interface.
As we show  below, for finite values of $\msqc$ a crossover takes place between these two limiting cases.

Both for $\msqc\to \infty$ and $\msqc\to 0$, 
we observe in Fig.~\ref{fig_PS} that the avalanche-size distribution has a cutoff 
for large avalanches
that is controlled by the system size $\Lmax$, and scales as $S_{\rm max}\sim L^{\dim+\zeta}$. 
In contrast, 
for uniformly driven interfaces   the avalanche size  is controlled by the driving spring: $\Savamax\sim1/m^{\dim+\zeta}$ (in the usual situation in which $m^{-1}<L$).
This difference is a consequence of the fact  
that in the steady state the system  moves uniformly on average.  As a consequence, 
since under localized driving  the driving point is     part of every avalanche, 
there are system-spanning avalanches. 
If we separate   avalanches of length $\lava<L$ 
from those with $\lava=L$ and plot the two separate distributions 
(dashed and dotted lines 
shown in Fig.~\ref{fig_PS}), we see that the ``bump'' observed at large sizes for the whole distribution comes from system-spanning avalanches.
If we consider the distribution of avalanche sizes restricted to the ensemble of avalanches with $\lava<\Lmax$, % strictly, 
as shown by the squares in the inset of Fig.\ \ref{fig_PS}, we observe a cutoff scaling as $\Savamax\sim\Lmax^{2.25}$ both in the soft and hard spring limit. 
Since in the uniformly driven case $\Lmax^{\dim+\zeta} \approx \Lmax^{2.25}$ is the average size of avalanches of length $\Lmax$, this   confirms that the roughness exponent $\zeta=1.25$ for both driving protocols, even  though metastable configurations in the tip-driven case are not flat on average as those of the uniformly driven case, but parabolic. This is the quenched Edwards-Wilkinson depinning universality class.

Now consider the scaling of the cutoff of system-spanning avalanches (circles in the inset of Fig.~\ref{fig_PS}), for which we obtain a
different behaviour in each limit. 
When $\msqc\to0$, the system moves rigidly\footnote{Actually, the points next to the tip moves slightly less than others but this is negligible.} some fixed distance
 $\Sava/\Lmax$,  and  $\Savamax\sim\Lmax^{\zetadep}$, thus system-spanning avalanches  have the same statistics as a particle ($d=0$) in an effective potential with characteristic scale $\Lmax^{\zetadep}$.
In contrast, when pulling the system with a stiff spring $\Savamax\sim\Lmax^{1+\zetadep}$, showing that system-spanning avalanches still behave as a 1-dimensional system.
{We leave a more detailed analysis of this dimensional crossover for future work.}

\begin{figure}
\includegraphics[width=0.95\linewidth,page=1]{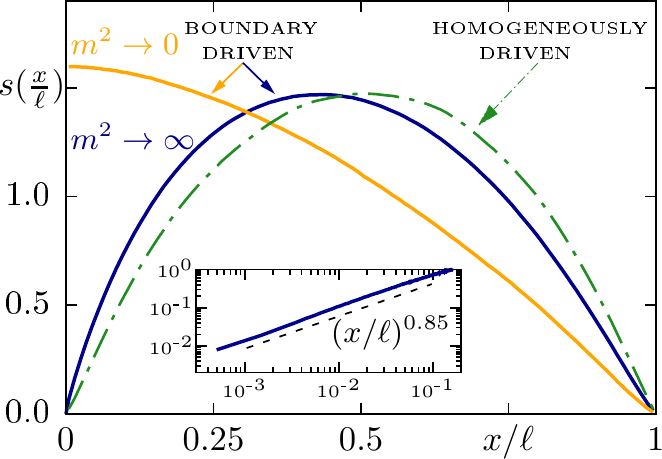}
\caption {\small (Color online) Reduced avalanche shape $s(x/\ell)=\Savaxaverage_{\lava}/\lava^\zetadep$ for uniformly driven (dashed blue line), tip driven with $\msqc\to\infty$, (solid red line) and tip driven with $\msqc\to0$, (solid brown line).
Inset: tip driven  $\msqc\to\infty$ case in log scale, showing $\sim x^{\theta}$ behaviour with $\theta=0.85$. 
}
\label{fig_sLreduce}
\end{figure}

The cases $\msqc \to 0$ and $\msqc \to \infty$ can also be  distinguished by the average spatial profile  $\Savaxaverage_\lava$ of the avalanches. 
For avalanches of the same length $\lava$ and points belonging to the avalanche $x<\lava$,
we verify $\Savaxaverage_\lava \sim \lava^{\zeta} s(x/\lava)$ for large enough $\lava$.
In Fig.~\ref{fig_sLreduce} we show for both limits the reduced shape $s(x/\lava),$ choosing $\lava\approx 2500$.
For comparison, the case of uniform driving is   included %in the figure 
($\msqc \Lmax = 100$). 
Within our numerical precision the average spatial profile for   uniform driving displays a form   indistinguishable from a parabola, vanishing linearly at the avalanche's extremes. To our knowledge this result has not been reported previously in the literature.

In the tip-driven case, the value of $\msqc$ has a strong influence on the avalanche shape.
When $\msqc\to0$, the tip %boundary 
can move freely during an avalanche,  the maximum displacement takes
place at or near the driven boundary, and  the average profile is half   the parabolic profile of the uniformly driven case described previously\footnote{A small difference is   observed at the maximum due to the fact that we are pulling on the edge and not on an inner site of the system.}. 
This is consistent with the observation of the avalanche size exponent $\tau\approx 1.11$ (Fig.~\ref{fig_PS}), corresponding to the uniformly driven case, Eq.\ (\ref{tau2}).
This  seems   natural since  this situation  is like   a localized constant-force driving $\sigma_1=\msqc w,$  as one may neglect the tip position $u_1$ compared to the driving position $w$. 
Localizing the driving when $\msqc\to0$ thus only   imposes the
starting point of the avalanche, but   does not change the physics as compared to a homogeneously driven system.

In contrast, when $\msqc\to\infty$ the above argument fails, and one must consider $\sigma_1=\msqc(w-u_1)$. 
Such a stiff driving imposes 
a constant displacement at the tip during the quasi-static dynamics, and     strongly restricts the tip displacement during an avalanche, 
resulting in an avalanche profile that     vanishes at this point. 
Interestingly, the %resulting 
corresponding   avalanche profile  has an asymmetry, in contrast to the symmetric shape for uniform driving.
Moreover, our %numerical 
results show that the avalanche profile for small $x$ starts  as $\sim x^\theta$, with $\theta \simeq 0.85$. 
We  have no clear understanding of the origin of this behaviour, and whether the exponent $\theta$ can be expressed in terms of the roughness exponent $\zeta$. 
To summarize: Although $\Savaxaverage_{\lava} \sim \lava^{\zeta}$ in all cases, with the same roughness exponent $\zeta=1.25$, the avalanche shapes have  distinct shape functions 
$s(x/\lava) = \Savaxaverage_{\lava}/\lava^\zetadep$.

\begin{figure}
%\centering
\includegraphics[width=0.95\linewidth,page=1]{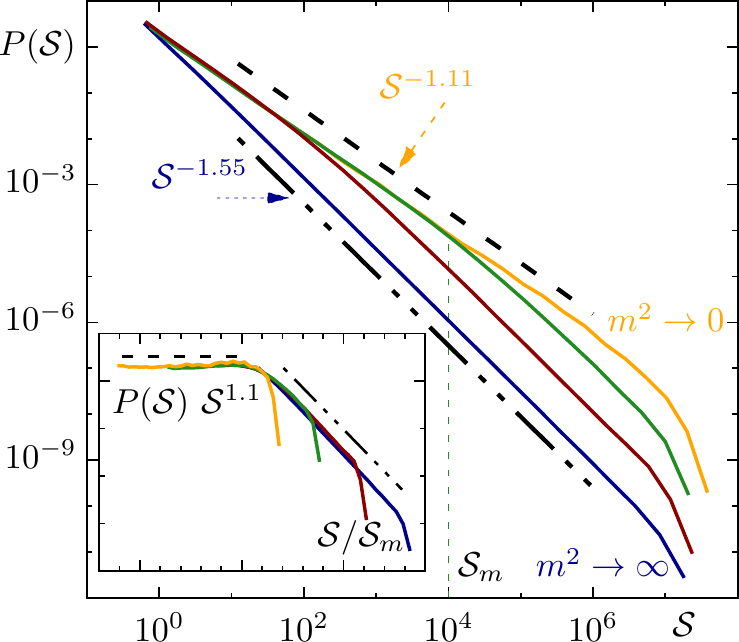}
\caption{
\small (Color online) Avalanche size distributions $\PS$ on a system of size $\Lmax=3980$, for different values of $\msqc$. The value of the crossover $\Scross$ is indicated on the green curve which corresponds to $\msqc=10^{-2}$. Inset: Scaled form using $\Scross\sim m^{-4.5}$.
}
\label{fig_PScross}
\end{figure}

We now discuss the case of  finite $\msqc$, and the crossover from 
$\msqc\to 0$ to $\msqc\to \infty$ for a  tip-driven interface.
For intermediate values of $\msqc$ we expect to see a crossover between the two limiting values of the exponent $\tau$  
(from $\tau\approx 1.11$ to $\tau\approx 1.55$ of Fig.~\ref{fig_PS}), and between the two limiting reduced avalanche  shape functions $s(y)$: From a parabolic shape with the maximum at zero to an asymmetric profile with the maximum at $x<\lava/2$,   see  Fig.~\ref{fig_sLreduce}. 
In Fig.~\ref{fig_PScross} we present the avalanche-size distribution $\PS$ 
for different values of $\msqc$ at  fixed system size $\Lmax=3980$.
Only avalanches   smaller than the total system size are considered ($\lava<\Lmax$).
As   in the extreme cases, for any value of $\msqc$ the avalanche-size cutoff is  controlled by the system size. 
Approximate power law decays for $P({\cal S})$ are   observed, but the exponent $\tau$ does not vary continuously from $\tau\approx 1.11$ to $\tau\approx 1.55$. Rather, it develops two power-law regimes separated by a characteristic $m$-dependent crossover scale $\Scross$: for $\Sava<\Scross$, $\tau\approx 1.11$, while for $\Sava>\Scross$, $\tau\approx 1.55$. 
As shown in the inset of Fig.~\ref{fig_PScross}, $\Scross \sim m^{-4.5}$. 
The limit $\msqc\to \infty$ is thus an attracting fixed point for the asymptotic behaviour at any finite $\msqc$. 

The crossover scaling can be understood as follows: For an avalanche of extension $\lava$, the maximum displacement
scales as  $\lava^\zetadep$, and the typical elastic force   as  $ c\lava^{\zetadep-1}$.
At the crossover scale, the displacement of the tip is of the
order of the maximum displacement $  \lava^\zetadep$, and
the force of the driving spring is 
$  m^2 \lava^\zetadep$. 
Balancing these two forces yields 
$\lcross\sim m^{-2}$. Hence, the crossover is expected at 
\begin{equation}
S_m^{\rm cross} \sim L_m^{1+\zeta} \sim m^{-2(1+\zeta)}.
\end{equation}
which gives $\Scross \sim m^ {-4.5}$
This scaling  matches   well our numerical results.
It should be exact, as indicated by the following 
argument: 
The total elastic force is 
$c \int {\rm d}  x \nabla^2 u(x)$
and the total driving force is  
$\int {\rm d} x\, m^2 \delta(x) [u(x)-w]$.
It can be proven that due to the statistical tilt symmetry $c$ and $m^2$ are not renormalized, validating the above argument. 
It is worth noting that for the uniformly driven system, a characteristic length scaling as $m^{-1}$ rather than $m^{-2}$ controls the avalanche-size  cutoff whenever $m^{-1}<L$. 

\begin{figure}
\includegraphics[width=0.95\linewidth,page=1]{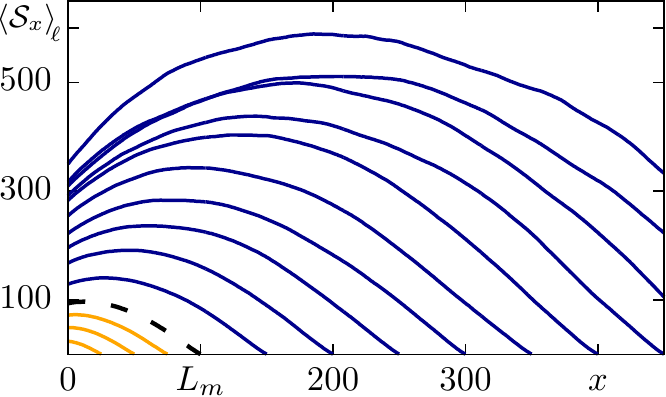}
\caption{\small (Color online) Average avalanche shape $\Savaxaverage_{\lava}$ for avalanches of different lengths $\lava$ %($\lava=10,30\dots170,190$) 
on a system of size $\Lmax=1584$ and a driving spring $\msqc=10^{-2}$ at the boundary.
The crossover avalanche length $\lcross$ is indicated.}
%The inset shows the scaling collapse by plotting the reduced shape $\Sshape (x/\lava)=\Savaxaverage \lava$ VER DE QUE PUEDE IR EN EL INSET Y COMO.
\label{fig_sLcross}
\end{figure}

To better understand this crossover, 
in Fig.~\ref{fig_sLcross} we plot the average spatial profile $\Savaxaverage_{\lava}$ of avalanches   
for different extensions $\lava$ using $\msq=10^{-2}$.
We observe that the avalanche shape has its maximum at the boundary for small avalanches,  whereas the maximum moves to increasing values of $x$ for increasing $\ell$.  
As indicated by the bold dashed line in Fig.~\ref{fig_sLcross}, this transition occurs when the extension $\ell$ reaches the crossover scale $L_m$ identified previously, i.e.\ for  avalanches of size  $\Scross \sim \lcross^{1+\zetadep}$, see Fig.~\ref{fig_PScross}. 

In addition to global avalanches, we can look at local jumps. Let us first consider a typical position $\phi$ \textit{within} the bulk of an avalanche which jumps a distance $\Savaphi \sim \lava^\zetadep$.
Using that    avalanche extensions scale as $P(\ell)\sim 1/\ell^{(\tau-1)(\dim+\zeta)-1}$, and equating $P(\ell){\rm d}\ell \simeq P({\cal S}_\phi){\rm d}{\cal S}_\phi $ we obtain that the local jump distribution satisfies

\begin{equation}
P(\Savaphi)\sim \Savaphi^{-\tau_{\phi}}\ ,\qquad  \tau_{\phi}=\tau + \frac{ (\tau -1)}{\zeta }
\ .\end{equation}
The value of $\tauS$ depends on the size $\Sava$ of the avalanche to which 
the displacement $\Savaphi$ belongs. 
Note that the exponent is distinct from the one for bulk driving given in \cite{DobrinevskiLeDoussalWiese2014a}; the difference is that here the point $\phi$ is inside the avalanche, since it is driven, whereas in \cite{DobrinevskiLeDoussalWiese2014a} it is an arbitrary point. 

\begin{figure}
\centering
\includegraphics[width=0.95\linewidth]{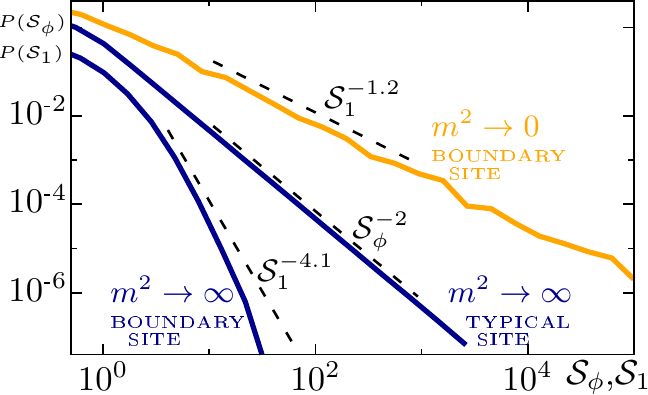}
\caption{
\small (Color online) Local avalanche size distributions for typical and boundary  sites, $P(\Sava_{\phi})$ and  $P(\Savatip)$ respectively. 
The power laws indicated are those 
given by Eqs. (\ref{pdesfi}) and (\ref{pdes1}). Note that for $m^2\to 0$  the boundary site is also a typical site. 
}
\label{fig_PS1}
\end{figure}

We have shown that avalanches with $\lava < \lcross < \Lmax$ have $\tauS = 2-2/(1+\zeta)$ while 
for $\lcross < \lava < \Lmax$ they have $\tauS = 2-1/(1+\zeta)$, 
both with $\zeta=1.25$, and  $\lcross \sim m^{-2}$ (see Fig.~\ref{fig_PScross}). 
Using these exponents we find
\begin{equation}
P(\Savaphi)\sim {\Savaphi}^{-\tau_\phi}
\begin{cases} \tau_\phi = 2-\frac{1}\zeta \approx 1.2& \mbox{if } \Savaphi^{1/\zeta}<\lcross<\Lmax \\ 
\tau_\phi = 2 & \mbox{if } \lcross<\Savaphi^{1/\zeta}<\Lmax \end{cases}.
\label{pdesfi}
\end{equation}
Interestingly,  %in $d=1$ considered here, 
local jumps driven by a hard spring 
are distributed with an exponent independent of $\zeta$. 
In Fig.~\ref{fig_PS1}, we verify these predictions by taking the limits $\msqc\to \infty$ and $\msqc\to 0$.
We also verified that a typical point belonging 
to a uniformly driven avalanche is   power-law distributed with an exponent $\tau_\phi=1.2$ (not shown).

For $\msqc\to 0$, the driven point with displacement $\Savatip$, is a typical site of the avalanche and its displacement is  thus distributed with an exponent $\tau_\phi=1.2$. 
However, for $\msq \to \infty$, the tip is not a typical point, and its displacement is  sensitive to the spatial avalanche profile near the border. 
Using that points near the boundary have  
$\Sava_x \sim \lava^{\zeta} (x/\lava)^{\theta} \sim \lava^{\zeta-\theta} x^{\theta}$ 
(see Fig. \ref{fig_sLreduce}),  
and that $P(\lava)\sim \lava^{1+\zeta}$ we get 
\begin{equation}
P(\Savatip)\sim \Savatip^{-\frac{2\zetadep-\theta}{\zetadep-\theta}}.
\label{pdes1}   
\end{equation}
Using $\theta=0.85$ and $\zetadep=1.25$ results in $P(\Sava_1)\sim \Sava_1^{-4.1}$.
The numerical results in Fig.~\ref{fig_PS1} seem to indicate an even steeper decay of $P(\Savatip)$ 
than the predicted power-law form. 
We believe that this behavior is due to rather strong finite-size effects, and that the exponent will converge to the correct asymptotic value for larger  $\lava$. However, at present the verification of this
statement is beyond our numerical capacities.

The statistics of the tip displacement is  accessible experimentally.
For example, when vortices are 
confined to a twin boundary of a superconductor and driven by a STM 
tip\cite{Shapira_2015}, the stress on the tip is $\sigma_1(t)=\msq[ w(t)-u_1]$, and the distribution 
of stress drops $\Delta \sigma_1 =\msqc \Sava_1$ can be measured by the driving device.

Key features of our results are consistent with recent exact
calculations   for the Brownian force model (BFM) driven at a  boundary
 \cite{DelormeLeDoussalWiese_2015}.
In this model, the disorder forces are Brownian walks,  and 
the exponent $\zeta$ takes the
value $4-d$, hence $\zeta=3$ for $d=1$. For $m^2=\infty$ the BFM has an avalanche-size exponent
$\tau=7/4$ in agreement with %our formula (1) 
Eq.~(\ref{tau1}),
setting $\zeta=3$. For $m^2 \to 0$ it yields $\tau=3/2$, 
the usual mean-field exponent %s 
for bulk driving. 
The crossover between the two scenarios occurs at   ${\cal S}_m \sim m^{-8}$, hence $L_m \sim m^{-2}$, in
  agreement with the above. The local
jump exponent $\tauphi$ takes the value $\tauphi=5/3$ for finite $m^2$, in agreement with Eq.~(\ref{pdesfi}) setting $\zeta=3$, and $d=1$.

\section{Conclusions}
In this Letter we studied the avalanche dynamics of an elastic line in a random medium driven at a point
by a spring of stiffness $\msq$ moving at constant velocity, 
and compared it with the well-known homogeneously driven case, 
where springs are attached to every point of the interface.
In both cases, universal scale-free power laws  with a large-size cutoff are observed in the distribution of avalanche sizes.
When driving the system homogeneously, 
the scale controlling the cutoff is given by the ratio between   
the elastic constants of the interface, $\cel$ and $\msq$, namely, $\lcross\sim\cel/m$, displaying criticality only in the limit $\msqc\to0$.
In contrast, when localizing the driving, the cutoff is controlled by the system size, and avalanches with an extent $\lava$ larger than $\lcross$  occur. 
This makes the locally driven elastic line \ a paradigm for driven self-organized critical systems.
Now $\lcross$ scales as $\sim(\cel/m)^2$ 
and becomes a crossover scale between two distinctive behaviours. 
We showed that small avalanches ($\lava<\lcross$) behave as in the homogeneously driven case with an exponent $\tau\approx 1.11$, 
while large avalanches ($\lava>\lcross$) present a new behaviour, with a higher avalanche exponent $\tau\approx 1.55$.

In these two regimes we measured the mean spatial shape of avalanches, a novel result for both homogeneously and locally driven elastic lines. 
They are distinct, changing from a seemingly universal symmetric parabolic shape to an asymmetric one, non-linearly growing at the driven point. 
Further work is needed to understand the origin of these shapes, and to clarify wether the characteristic exponents are related to the roughness exponent $\zetadep$ of the interface. 
Interestingly, both regimes have the same value $\zetadep=1.25$, corresponding to the quenched Edwards-Wilkinson depinning universality class.
We also measured the local distribution of jumps at different points of the interface,
which exhibit new critical exponents in each regime. 
We consistently find a small slope at small jump lengths crossing over to a much steeper value at larger jump lengths.

This motivates to search for a similar crossover in experiments and suggests 
new measurements. For instance, vortices driven by  an STM tip 
show a marked hysteresis \cite{Shapira_2015} a signature of the non-equilibrium effects
studied here, as well as a crossover in the jump-size distributions. We suggest to simultaneously  
measure jumps on the far side of the sample  to distinguish   system-spanning
avalanches from smaller ones, a distinction which proved to be important in our analysis. 

Finally, let us stress that most of the present results, and in particular scaling relations, can be generalized to an arbitrary spatial dimension $d$ (with an appropriate definition of the boundary driving mechanism). 
In particular, the case $d=2$, with appropriate modifications to the elastic kernel,  may be applied to the study of stick-slip motion observed in friction experiments and its relation to the dynamics of edge-driven tectonic faults \cite{Rubinstein_2011}. 

\acknowledgments
A. B. K., E. A. J, and L. E. A acknowledge partial support from Projects PIP11220120100250CO (CONICET) and PICT2011-1537 (Argentina). 
P. L. D. and K. J. W. acknowledge support from PSL grant ANR-10-IDEX-0001-02-PSL and thank KITP for hospitality and support in part by
the NSF under Grant No. NSF PHY11-25915. 

\bibliographystyle{eplbib}
\bibliography{tipdriven}{}

\end{document}